\documentclass[11pt]{article}
\usepackage{authblk}

\usepackage{mathtools}
\usepackage{amsmath}
\usepackage{amsfonts}
\usepackage{listings}
\usepackage{graphicx}
\usepackage{longtable}
\usepackage[parfill]{parskip}
\usepackage{bbm}
\usepackage{float}
\usepackage{adjustbox}
\usepackage{algorithm}
\usepackage{algorithmic}
\usepackage{caption}
\usepackage{hyperref}
\usepackage{multirow}
\usepackage{rotating}
\usepackage{newtxtext, makeidx, multicol, footmisc}
\usepackage{subfig}

\usepackage{booktabs}

\newcommand{\E}{\mathrm{E}}

\usepackage{xcolor}

\usepackage{lineno}
\title{A clusterability test for directed graphs}

\author[1]{Mario R. Guarracino}
\author[2]{Pierre Miasnikof}
\author[3,4]{Alexander Y. Shestopaloff \thanks{corresponding author: a.shestopaloff@qmul.ac.uk} }
\author[1]{Houyem Demni}
\author[5]{Cristi\'an Bravo}
\author[6]{Yuri Lawryshyn}
\affil[1]{Università degli Studi di Cassino e del Lazio Meridionale, Cassino, Italy}
\affil[2]{Universit\'e Laval, Qu\'ebec, QC, Canada}
\affil[3]{Queen Mary University of London, London, United Kingdom}
\affil[4]{Memorial University of Newfoundland, St-John's, NL, Canada}
\affil[5]{University of Western Ontario, London, ON, Canada} 
\affil[6]{University of Toronto, Toronto, ON, Canada}

\begin{document}
\maketitle

\begin{abstract}
In this article, we extend a statistical test of graph clusterability, the $\delta$ test, to directed graphs with no self loops. The $\delta$ test, originally designed for undirected graphs, is based on the premise that graphs with a clustered structure display a mean local density that is statistically higher than the graph's global density. We posit that graphs that do not meet this necessary (but not sufficient) condition for clusterability can be considered unsuited to clustering. In such cases, vertex clusters do not offer a meaningful summary of the broader graph. Additionally in this study,  we aim to determine the optimal sample size (number of neighborhoods). Our test, designed for the analysis of large networks, is based on sampling subsets of neighborhoods/nodes. It is designed for cases where computing the density of every node's neighborhood is infeasible. Our results show that the $\delta$ test performs very well, even with very small samples of neighborhoods ($1\%$). It accurately detects unclusterable graphs and is also shown to be robust to departures from the underlying assumptions of the $t$ test.
\end{abstract}

\section{Introduction}
In this article, we extend a statistical test for clusterability, the $\delta$ test \cite{PMEtAlLion19,PMEtAlPwr2023}, to unweighted directed graphs with no self loops. The $\delta$ test is based on the observation that clusterable graphs, graphs whose structure can be meaningfully summarized by clusters of vertices, will display high neighborhood-level densities, on average. It is designed to detect heterogeneity in connections, the presence of statistically significant pockets of high density. It can be used to answer two separate but related questions: {\it ``does graph $G$ have a high local connectivity pattern"} (i.e.,  meet the necessary condition) and {\it ``does graph $G_1$ have a higher local connectivity pattern than graph $G_2$''} \cite{PMEtAlLion19,PMEtAlPwr2023}. The $\delta$ test offers two major advantages. First, it uses samples of nodes to answer these two questions, which makes it suitable for the study of large networks. Second, it offers formal statistical answers with significance levels.

Graph clustering, also referred to as network community detection, is a pivotal task in the study of networks. In fact, it has even been described as a pivotal question in network analysis \cite{LiudaSynthetic2019}. Clustering algorithms will always group data into clusters, regardless of the data set's structure. However, such techniques should be used with care. If the data does not have a clustered structure, if it cannot be meaningfully summarized by grouping it into subsets of similar items, the clustering exercise is futile. Not only is it a waste of time, it leads to misleading conclusions. For this reason, clusterability tests play a key role in clustering analyses. They identify data sets (graphs in this specific case) that possess the prerequisite (necessary) conditions for being clusterable. To emphasize this fact, we cite Adolfsson et al.~\cite{AdolfssonEtAl2019} who stated: {\it``(...) an even more fundamental issue than algorithm selection is when clustering should, or should not, be applied''}, in their work on clustering in $\mathbb{R}^n$.

Currently, there is no formal generally accepted definition of (vertex) clusters or communities in undirected graphs. However, there is a quasi-universal agreement that clusters are formed by densely connected subsets of nodes that are sparsely connected to the remaining graph (e.g., \cite{Schaeffer2007,FortunatoLong2010,YangLesko2012,modWAW2016,guideFortunato16,EuroComb2017,PMEtAlWAW18,PMEtAlOUP20}). The original delta test is based on this quasi-universal agreement \cite{PMEtAlLion19,PMEtAlPwr2023}.

In the case of directed graphs, the definition of clusters is even more ambiguous. Many authors analyzing directed graphs simply ignore edge directions (e.g., \cite{NewGirvOrig2004,FortunatoLong2010,modWAW2016,EuroComb2017}). Case in point, in his seminal article, Fortunato stated {\it ``(...) it is customary to neglect the directedness of the hyperlinks and to consider the graph as undirected, for the purpose of community detection''} \cite{FortunatoLong2010}. However, many authors disagree with this simplistic treatment of directed graphs (e.g., \cite{directedCommunity2008,Mallias2013}).

Nevertheless, the characterization of clusters as densely connected subsets of nodes that are sparsely connected to the remaining graph seems to be accepted even by authors who do not agree with the na\"ive treatment of directed graphs as undirected ones. For example, Malliaros and Vazirgiannis \cite{Mallias2013} began by stating{\it ``(...) the most common way to dealing with edge directionality during the clustering task, is simply to ignore it. (...) However, in many cases, this simplistic technique would not be satisfactory (...)''}. Nevertheless, these same authors later described clusters as {\it ``(...) modules with dense connections between the nodes of the same cluster but sparser connections between different clusters''}.

The extension of our test to directed graphs is justified by this consensus agreement that clusters are formed by densely connected subsets of nodes that have only sparse connections to the remaining graph. In the case of directed graphs, we again posit that a local density that is, on average, significantly higher than the graph's global density is a prerequisite condition for the presence of meaningful clusters. On the basis of this postulate, we devise a test that samples local neighborhood densities and compares them statistically to the graph's overall density. Our test is designed to use only a small subset of all neighborhoods, which makes it very well suited as a first step in the analysis of large networks. It prevents wasted time on and misleading results from clustering graphs that are evidently unsuited for clustering. Moreover, because it only uses a small portion of the graph, it is feasible to apply it to larger data sets. 

The remainder of this article is organized as follows. After a very brief review of previous work on clusterability testing, we introduce our modification to the $\delta$ test. We then examine the performance of our test on synthetic graphs with known structure and real-world networks as well. We also investigate the statistical power of our test, its sensitivity to sample size. Our results show that the $\delta$ test performs very well. It accurately detects unclusterable graphs and is also shown to be robust to departures from the underlying assumptions of the $t$ test, even with a very small sample of nodes/neighborhoods.

\section{Previous work}
As in earlier work on the $\delta$ test, the extension presented here is heavily inspired by the statistical tests presented by Gao and Lafferty \cite{GaoLaff2017Prob,GaoLaff2017Stat}, as well as the follow-up work of Gao and Ma \cite{Gao2018}.  However, as shown in Miasnikof et al.~\cite{PMEtAlLion19}, the tests of Gao and Lafferty rely on restrictive assumptions, are not suited for graph comparisons and are unresponsive to graph structure.  Indeed, at the core, these tests focus on the presence of very restrictive structures (triangles) as indicators of a clustered structure. Such an approach was deemed too restrictive by Yin et al.~\cite{Yin2018}, who highlight the need to examine higher order cliques. Meanwhile, the work of Gao and Ma \cite{Gao2018} also imposes restrictive assumptions and is exclusively focused on undirected graphs.

In concluding this very brief literature review, we wish to highlight the fact that testing graphs for various properties is not new. It was initially introduced by Goldreich et al.~\cite{GoldEtAl1998}, in the late 1990s. These authors' seminal work introduced the practice of sampling vertices and testing for specific properties, which spawned a series of other articles on the same topic (e.g., \cite{AdriaensApers2021,PPtest2021,EstTriangles2021}). Additionally, we note that clusterability remains a topic of discussion in the literature (e.g., \cite{sparseClust2023}).

\section{Methods}
As described earlier, our test is based on a random sampling procedure. We begin by sampling a set of `$\ell$' ($ \ell \ll N$) nodes and computing the density of the induced subgraph formed by each sampled node's neighborhood. We then compare the mean density of the induced subgraphs to the graph's global density, statistically. In other words, we sample local clustering coefficients \cite{NewmanClustCoeff2018}, take their mean and compare it statistically to the graph's global density. Our test rests on the postulate that a necessary condition for a graph to be clusterable is that it possess heterogeneous local connection patterns. Indeed, a clusterable graph is composed of pockets of highly connected nodes with sparse connections to the remaining graph. Therefore, it must display highly non-uniform local neighborhood densities, with a non-trivial amount of pockets of higher (than global) density. 

Our statistical hypothesis, based on our postulate of heterogeneous densities, is that a clusterable graph will have a sample mean of local densities that is significantly higher than the graph's global density figure. It is very important to note here that a higher mean local density is a necessary but not sufficient condition for clusterability. Alternately, an unclusterable graph will have uniform densities across the graph. Statistically, this uniformity translates to a local density mean that is indistinguishable from the global density figure.

\subsection{Statistical hypotheses}
Under the null hypothesis the mean local density is statistically indistinguishable from the graph's global density. In cases where the null hypothesis is not rejected, we can conclude the graph is not clusterable, since it does not meet the necessary conditions. Indeed, in such cases, we statistically verify the absence of a significant number of pockets of high connectivity that are indicative of a possible clustered structure.

In the alternate case, when we reject the null, local density is significantly higher than global density, the graph meets the necessary condition for clusterability. The presence of a significant number of highly connected pockets of nodes that are sparsely connected to the remaining graph is confirmed. Here, it is very important to highlight that rejecting the null does not offer a guarantee the graph is indeed clusterable. It simply means the graph meets the necessary condition for being clusterable.

We also extend our test to determine if a given graph is more clusterable than another. The question here is to determine if a graph $G$ is composed of a stronger clustered structure than graph $H$. In this case, the null hypothesis is that the mean local densities in both graphs are equal. Under the alternate hypothesis, the local density in graph $G$ is significantly greater than the one in graph $H$.

\subsection{Probabilistic interpretation}
As mentioned throughout this document, our goal is to assess the statistical significance of the difference between mean local and global densities. However, in order to justify such a statistical significance test, we must lay the probabilistic foundations that make such a test possible. Just as in our previous work \cite{PMEtAlLion19,PMEtAlPwr2023}, we highlight the fact graph and neighborhood induced subgraph densities can be interpreted as the probability that two vertices are connected by an edge. In this article, however, we extend this interpretation by ignoring edge direction. We interpret density as the probability that two vertices are connected by an edge in either direction.

Indeed, a graph's global density can be understood as the probability that two arbitrarily selected vertices are connected by an edge in either direction. Similarly, at the neighborhood level, local density can also be interpreted as a probability. It can be understood as the conditional probability that two nodes are connected in either direction, given they are both in the same induced subgraph formed by their common neighborhood. The equations below present a mathematical description of this probabilistic interpretation. Equation~\ref{uncondprob} shows the computation of a graph's global density, which we denote as $\mathcal{K}$. The set of edges is denoted by the usual $E$ and the set of vertices by $V$.  
\begin{eqnarray}
	\mathcal{K} &=& P(e_{ij} \lor e_{ji}) = \frac{\vert E \vert}{\vert V \vert \times (\vert V \vert-1)}  \label{uncondprob}\\
	\nonumber\\
	\kappa_{\tilde{\nu}} &=& P(e_{ij} \lor e_{ji} \vert \nu_i = \nu_j = \tilde{\nu} ) = \frac{\vert e \vert}{n \times (n-1)} \label{condprob}
\end{eqnarray}
In Equation~\ref{condprob}, we compute the neighborhood density for an arbitrary neighborhood $\tilde{\nu}$ containing $n$ vertices. The probability two arbitrary nodes $i$ and $j$ with a neighborhood $\nu_i = \tilde{\nu}$ and $\nu_j = \tilde{\nu}$ respectively are connected in the induced subgraph formed by this neighborhood is given by the ratio of the total number of edges in this subgraph ($\vert e \vert$) over the total number of possible connections.

Our statistical test is grounded in this probabilistic interpretation. We posit that a clusterable graph will have a non-trivial amount of densely connected neighborhoods. Under our postulate, vertices of a clusterable graph have a higher probability of sharing an edge with vertices having common neighbors than with those with which they don't. In the average case, we expect that vertices sampled from a given neighborhood will have a greater connection probability than vertices that are arbitrarily sampled across the graph, if the graph is clusterable.

Finally, we note that the mean local density $\bar{\kappa}$ is the empirical mean probability that two nodes in the same neighborhood are joined by an edge in either direction. Meanwhile, as just mentioned, the density $\mathcal{K}$ is a graph constant representing the probability two nodes are connected, regardless of their neighborhood membership. As a result, the probability distribution of the difference between $\bar{\kappa}$ and $\mathcal{K}$ can be approximated by a Gaussian (or Student's $t$) distribution, on the basis of the Central Limit Theorem (CLT) \cite{WeissCLT}. In this specific instance, we approximate the Gaussian with a Student $t$ distribution, since the population standard deviation is estimated from the data itself \cite{WeissStudent}. 

\subsection{Sampling procedure}
\begin{figure}[]
\centering
\includegraphics[width = 0.999\textwidth]{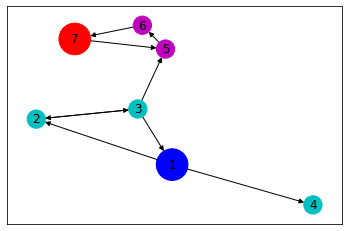}
\caption{Sampling example}
\label{ex}
\end{figure}

We use Figure~\ref{ex} to illustrate our sampling procedure: 
\begin{itemize}
\item In this example, we begin by randomly selecting nodes $1$ (blue) and $7$ (red).
\item We then compute the densities of each randomly selected node's neighborhood induced subgraph.
\item One subgraph is formed by the three neighbors of node $1$. It contains nodes $2,3$ and $4$ (i.e., $n_1 = 3$).
\item It has two (directed) edges, $(2,3)$ and $(3,2)$, node $4$ is not connected to the other two nodes (i.e., $\vert e_1 \vert = 2$).
\item This induced subgraph formed by the neighbors of node $1$ has density:
\[
\kappa_1 = \frac{\vert e_1 \vert}{n_1 \times \left( n_1 -1 \right)} =\frac{2}{3 \times 2} = 0.5 \, .
\]
\item The second subgraph is formed by the neighbors of node $7$.
\item It has two nodes ($5$ and $6$) and one edge $(5,6)$ (i.e., $n_2 = 2, \, \vert e_2 \vert = 1$).
\item Its density is
\[
\kappa_2 = \frac{\vert e_2 \vert}{n_2 \times \left( n_2 -1 \right)} = \frac{1}{2 \times 1} = 0.5 \, .
\]
\item The mean local density is
\[
\bar{\kappa} = \frac{1}{2} \left( \kappa_1 + \kappa_2  \right)= 0.5 \, .
\]
\end{itemize} 
Meanwhile, we also note the global density of this graph formed by seven vertices and nine edges is
\[
\mathcal{K} = \frac{\vert E \vert}{\vert V \vert \times \left( \vert V \vert - 1 \right)} = \frac{9}{7 \times 6} \approx 0.21 \, .
\]

\subsection{Testing algorithm} \label{algo}
As mentioned in the previous section, our $\delta$ test statistic can be used to answer two related but distinct questions:
\begin{enumerate}
	\item Does a graph display high local density?
	\item Does graph $G_1$ have a higher local density than graph $G_2$?
\end{enumerate}
In the first question, we ask whether a graph meets the necessary condition to have a clustered structure. In the second, we ask if one graph meets this condition more strongly than another. These questions can be answered by following these steps:
\begin{itemize}
\item Sample $\ell = \lfloor s \times \vert V \vert \rfloor$ vertices and extract the $\ell$ induced subgraphs formed by the neighborhood of each sampled node ($s$ is the percentage of nodes sampled whose neighborhoods' densities are computed in the next step);
\item Compute the local densities $\kappa_i = \frac{\vert E_{i} \vert}{0.5 \times n_i(n_i - 1)}$ for each of the $\ell$ subgraphs ($n_i$ is the number of nodes in the $i$-th subgraph);
\item Compute graph density $\mathcal{K} = \frac{ \vert E \vert}{\vert V \vert(\vert V \vert-1)}$, for the graph $G=(V,E)$;
\item Compute the mean of the local densities: $\bar{\kappa} = \frac{1}{\ell} \sum_{i=1}^\ell \kappa_i$;
\item Normalize the mean and obtain the test statistic:	$\delta = \frac{\bar{\kappa}}{\mathcal{K}} - 1$;
\item Under the null, the test statistic $\delta$ follows a Gaussian distribution centered at 0, which is approximated by a Student's $t$ distribution (of $\ell -1$ degrees of freedom);
\item Under the alternative hypothesis, we expect the $\delta$ statistics to still follow a Gaussian distribution, but one that is centered about about a point that is greater than zero;
\item In the case of a single-graph test, perform a one-tailed $t$ test; the null hypothesis is $\E(\delta) = 0$, the alternative is $\E(\delta) > 0$; 
\item In the case of a two graph test, perform a two sample (unpaired with unequal variance) one-tailed $t$ test on the difference $d_{GH} = \delta_G - \delta_H$. In this case, the null hypothesis is $\E(\delta_{G}) = \E(\delta_{H}) \Leftrightarrow E(d_{GH}) = 0$, the alternative is $\E(\delta_{G}) > \E(\delta_{H}) \Leftrightarrow E(d_{GH}) > 0$.
\end{itemize}

\section{Test validation and sample size determination}
In order to empirically assess the probability of rejecting the null hypothesis, we apply our test to graphs of various structures. We repeat our sampling and significance testing algorithm 500 times, on several synthetic graphs. Our goal is to estimate the accuracy, the false positive rates (type I error) and statistical power (type II error) of the $\delta$ test. We use graphs with a known structure and a-priori answer to the (un)clusterability question. 

Because our test only uses a small portion of all nodes, we also examine the sensitivity of rejection probabilities to sample size. In order to determine optimal sampling size, we repeat our 500 test iteration with samples of 10\%, 1\% and 0.5\% of nodes, using synthetic graphs whose (un)clusterability is known in advance. Once we have determined the optimal sampling size, we apply our empirical work with a test of practical applicability and scalability, using real-world networks. We end our tests with several synthetic and real-world two graph tests.

\subsection{Test graphs}
As mentioned earlier, we begin our tests with several synthetic graphs. These synthetic graphs offer a-priori known cluterability (or lack of). This knowledge allows us to validate our test and determine optimal sampling size. All our synthetic test cases were generated using the Python NetworkX library \cite{NetworkX2008}. 

In the case of real-world networks, we know that like virtually all curated publicly available research data sets, they all display a clustered structure. A summary description of each test case is provided below. We use these networks to illustrate the test's practical applicability and scalability to large networks.

\subsubsection{Synthetic graphs} \label{data_descript}
No clustered structure
\begin{itemize}
\item CM: A configuration model graph \cite{ConfigMod2003,NewmanConfigMod2018} with power law in/out degree sequences of exponent $3.5$ and $\vert V \vert = 7,000$
\item ER: An Erd\"os-R\'{e}nyi-Gilbert random graph \cite{EROrig,Gilbert1959} with directed edge probability of $0.333$ and $\vert V \vert = 7,000$
\end{itemize}

Clustered structure
\begin{itemize}
\item CC: A directed connected caveman graph \cite{WeissCave,NetworkXCC} with $140$ cliques of $50$ nodes each, for a total of $\vert V \vert = 7,000$
\item SBM: A directed stochastic block model graph with N nodes, M vertices, C communities containing 80 to 120 nodes each, with a mean inter-block edge probability of 0.3 and mean intra-block edge probability of 0.75 and $\vert V \vert = 6,910$ \cite{SBMOrig83,NetworkXSBM} 
\end{itemize}

\subsubsection{Note on configuration model graphs}
In this work, we use a (directed) CM graph with power law in/out degree sequences of exponent of $3.5$. However, previous experiments on the $\delta$ test, in the context of undirected graphs, were conducted on CM graphs with power law exponents of $3$ \cite{PMEtAlLion19,PMEtAlPwr2023}. Because the variance of power laws are only defined when the exponent is greater than $3$ \cite{NewmanConfigMod2018}, these tests served as ``stress test'' scenarios. The goal was to assess the robustness of the test, under departures from the assumptions underlying the $t$ test. Naturally, since these tests were conducted on finite samples, a standard error was computable, although high.

In the directed case, however, the undefined (i.e., asymptotically  infinite, finite but elevated in finite samples) variance of the power laws of exponent 3 inflates the clustering coefficient \cite{NewmanConfigMod2018} and pushes the $\delta$ test to misclassify these CM graphs as ``possibly clusterable'' (i.e., meeting the necessary conditions for clusterability) more often than expected under the null. Although the CM graphs typically do not have a clustered structured \cite{Hofstad2016}, our experiments reveal that under certain circumstances, a non-trivial amount of dense neighborhoods are indeed observed. With an exponent of $3.5$, which yields a theoretically finite variance, we observe that our $\delta$ statistic still has a non-stationary distribution. We posit that this distributional instability is the result of variance that is dependent on sample size. Nevertheless, the test remains robust, in spite of this significant departure from the null and alternate models. 

Unfortunately, it is difficult to explain these test results. However, we highlight the fact that very little is known about in/out degree distributions in directed graphs. For example, Newman stated in 2018: \textit{``In practice, the joint in/out degree distribution of directed networks has rarely been measured or studied, so there is relatively little data on it. (...) For the moment, however, this is an area awaiting more thorough exploration''} \cite{NewmanPwrLaw2018}. We also note that the same monograph describes the link between the second moment of the power law degree sequence and the (mean local) clustering coefficient \cite{NewmanPwrLaw2018}. Therefore, we can only suspect that the elevated variance observed in the undirected case is compounded in the directed case.

The case of the configuration model with high local clustering coefficients calls attention to a very important challenge in community detection, the very definition of vertex clusters. While vertex clusters or communities are arguably formed by densely connected subsets of nodes, we have observed that these subsets can be the product of a purely random assignment. This apparent paradox further underlines the necessity of combining both domain context specific expertise and algorithmic tools. Although debatably anecdotal, this example highlights the failure of a purely mathematical general approach.

\subsubsection{Real-world networks}
In order to illustrate the value of our test, we also apply it to large real-world networks. These two networks are known to have a clustered structure. They are also selected for their considerable sizes.
\begin{itemize}
\item AMAZON:  Amazon product co-purchasing network (``Amazon product co-purchasing network from May 05 2003''), $\vert V \vert = 410,236$ and $\vert E \vert = 3,356,824$ \cite{LeskovecRW2007} 
\item WIKI: ``Web graph of Wikipedia hyperlinks'' $\vert V \vert = 1,791,489$, $\vert E \vert = 28,511,807$ \cite{KlymkoRW2014,YinWiki2017}
\end{itemize}

\section{Empirical experiments}
We begin our experiments with synthetic graphs whose clusterability is known a priori. The goal of these experiments is to validate the test and determine the optimal sampling size. With the test validated and the optimal sampling size, we illustrate the value of our test on large real-world networks. We also illustrate the test in the two graph context.

\subsection{Single graph hypothesis tests, distribution of $\delta$ statistic and sample size determination}
In this section we show (single sample) hypothesis test experiments designed to study test results and the distribution of our $\delta$, with respect to sample size (percentage of nodes, '$s$'). In Table~\ref{exp500}, we show the number of rejections of the null hypothesis for hypothesis tests of the $\delta$ statistic. Each row represents results for tests with different subsets of nodes. More specifically, the first column shows the graph name, the second the percentage of nodes samples at each iteration, the third column shows the number of times the null hypothesis was rejected, the fourth column shows that same number expressed as a fraction of the number of trials. All experiments were repeated 500 times. 
\begin{table}[H]
\centering
\caption{Null hypothesis rejection} \label{exp500}
\begin{tabular}{lcccc}
\toprule
& Graph  & Prop Sample & Num rejects & Prop reject \\
\midrule
\multirow{3}{*}{No clusters} & CM     & 0.005      & 0            & 0     \\
& CM     & 0.01       & 0            & 0     \\
& CM     & 0.1        & 2          & 0.004      \\
\midrule
\multirow{3}{*}{No clusters} & ER.333 & 0.005      & 17        & 0.034     \\
& ER.333 & 0.01       & 19          & 0.038     \\
& ER.333 & 0.1        & 9           & 0.018     \\
\midrule
\multirow{3}{*}{Clusters}    & CC     & 0.005      & 428     & 0.856     \\
& CC     & 0.01       & 493         & 0.986     \\
& CC     & 0.1        & 500         & 1         \\
\midrule
\multirow{3}{*}{Clusters}    & SBM    & 0.005      & 499    & 0.998     \\
& SBM    & 0.01       & 500         & 1         \\
& SBM    & 0.1        & 500         & 1        \\
\bottomrule
\end{tabular}
\end{table}

\begin{figure}[H]
\centering
\subfloat[ER, 0.5\% of nodes]{ 
\includegraphics[width = 0.33\textwidth]{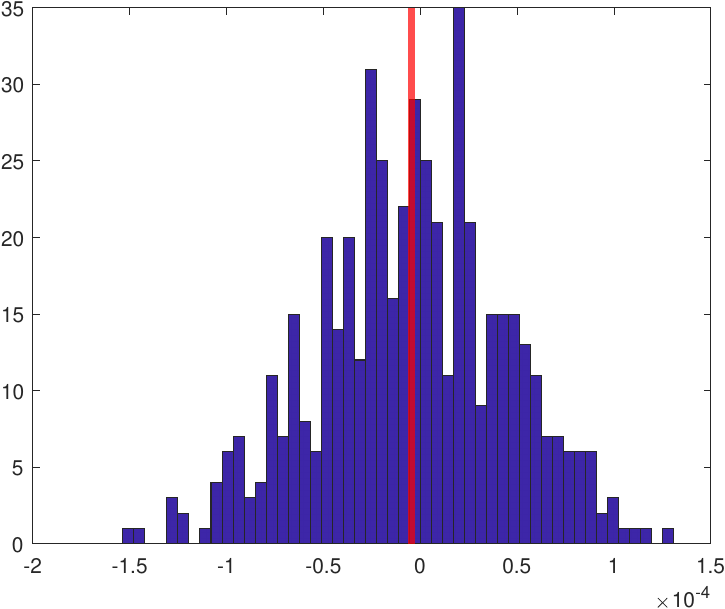} }
\subfloat[ER, 1\% of nodes]{ \includegraphics[width = 0.33\textwidth]{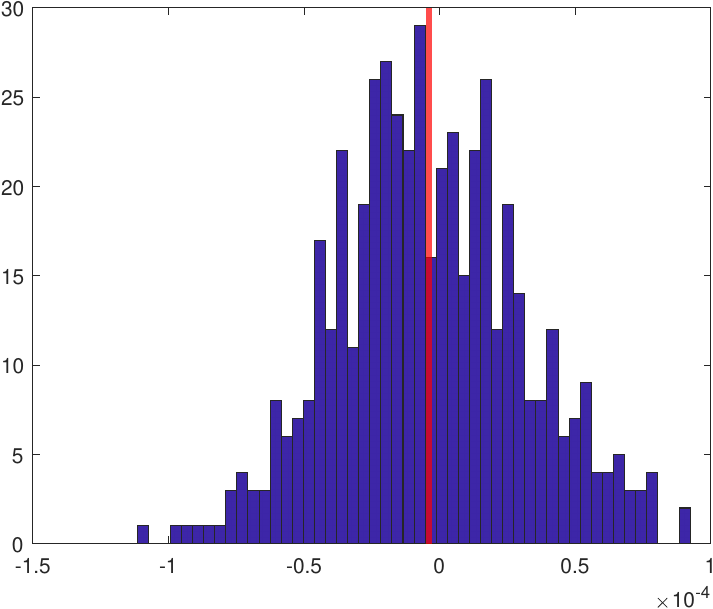} } 
\subfloat[ER, 10\% of nodes]{ \includegraphics[width = 0.33\textwidth]{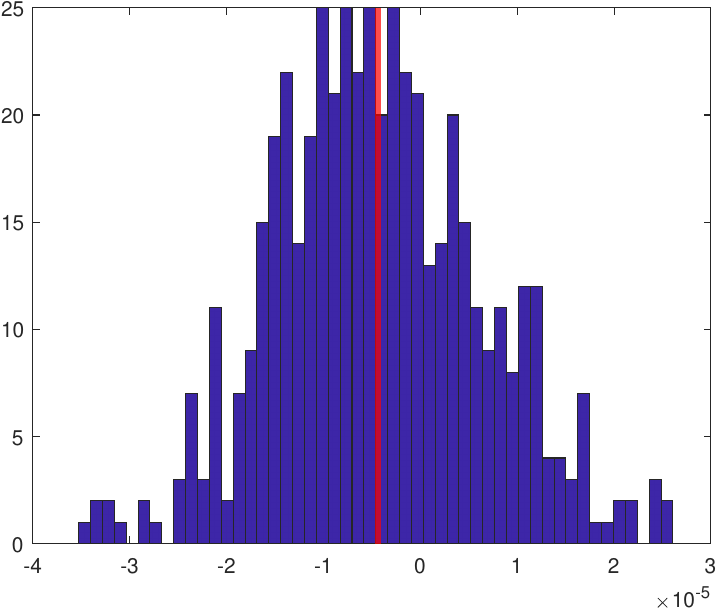} }
\\
\subfloat[CM, 0.5\% of nodes]{ \includegraphics[width = 0.33\textwidth]{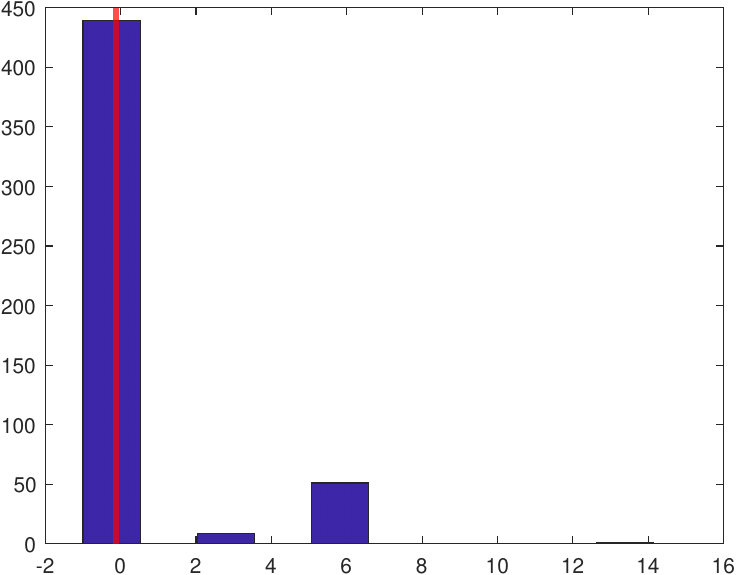} } 
\subfloat[CM, 1\% of nodes]{ \includegraphics[width = 0.33\textwidth]{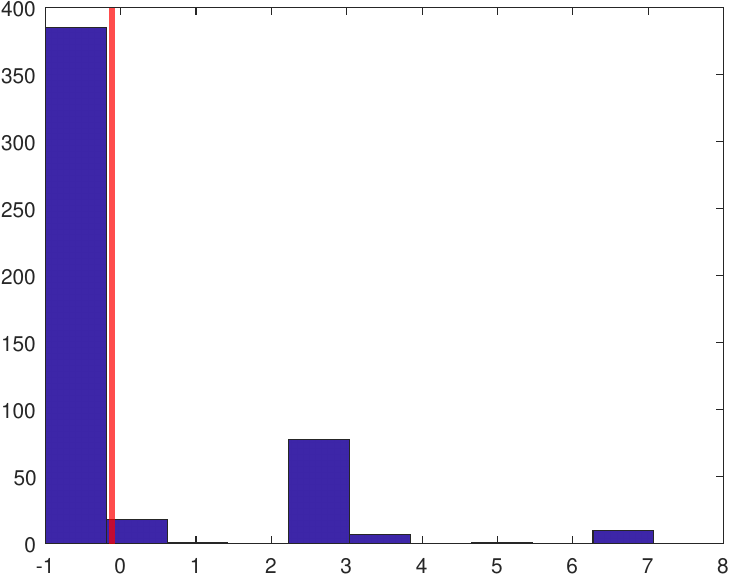} } 
\subfloat[CM, 10\% of nodes]{ \includegraphics[width = 0.33\textwidth]{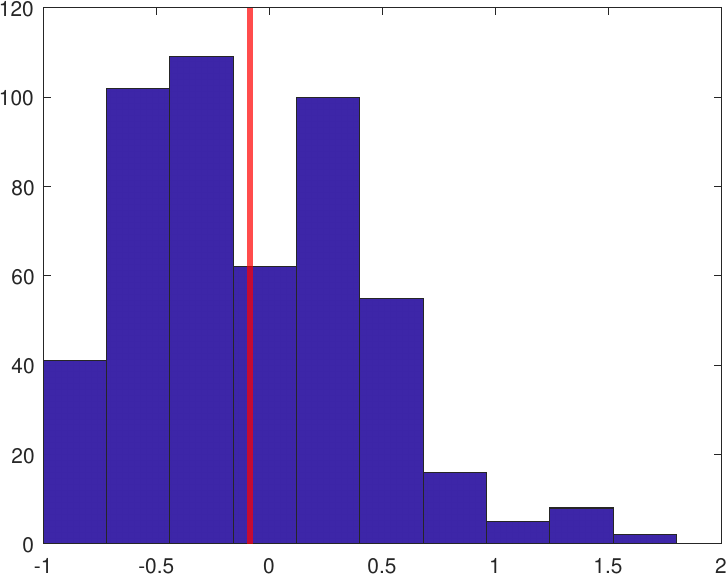} }
\caption{Distribution of $\delta$ statistic, 500 experiments, unclusterable graphs} \label{Synthhists1}
\end{figure}
\begin{figure}[H]
\centering
\subfloat[SBM, 0.5\% of nodes]{ \includegraphics[width = 0.33\textwidth]{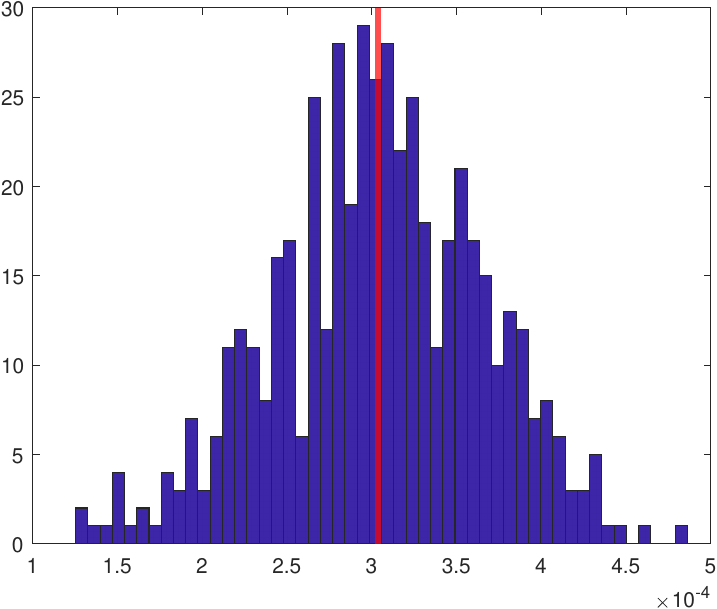} }
\subfloat[SBM, 1\% of nodes]{ \includegraphics[width = 0.33\textwidth]{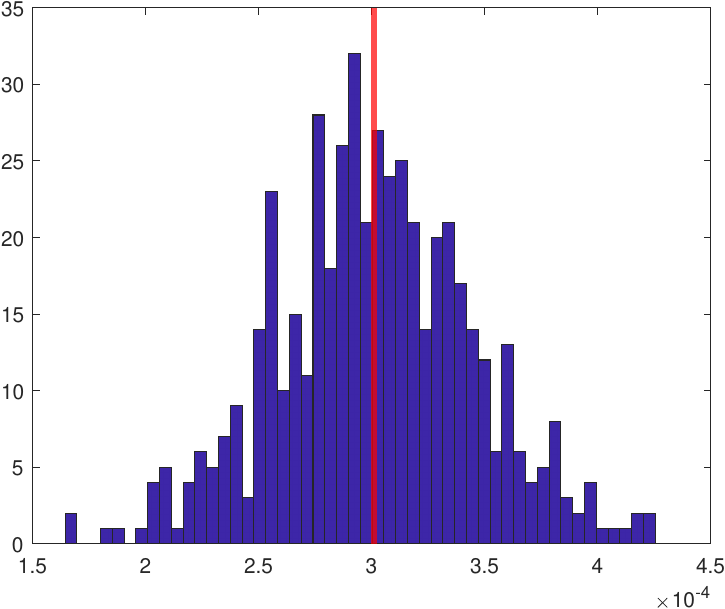} } 
\subfloat[SBM, 10\% of nodes]{ \includegraphics[width = 0.33\textwidth]{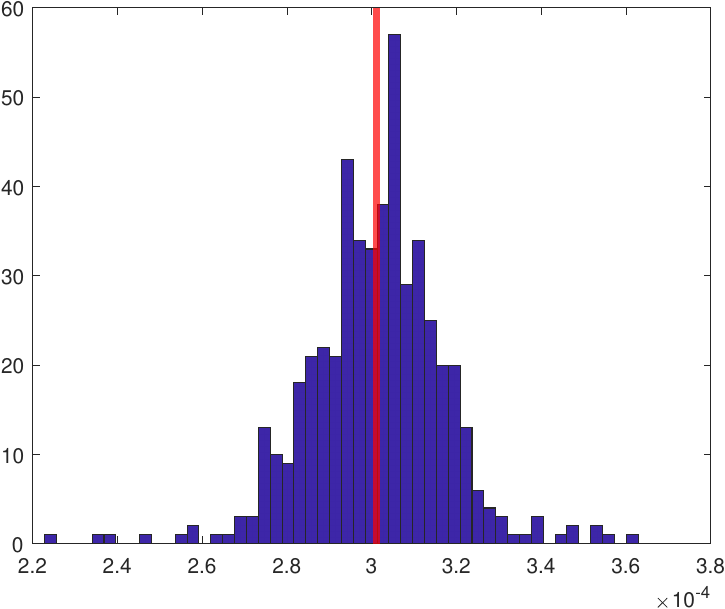} }
\\
\subfloat[CC, 0.5\% of nodes]{ \includegraphics[width = 0.33\textwidth]{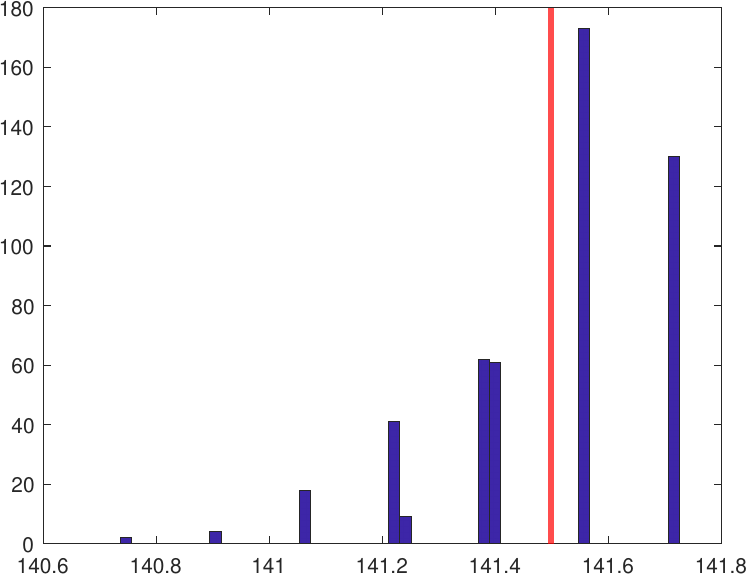} }
\subfloat[CC, 1\% of nodes]{ \includegraphics[width = 0.33\textwidth]{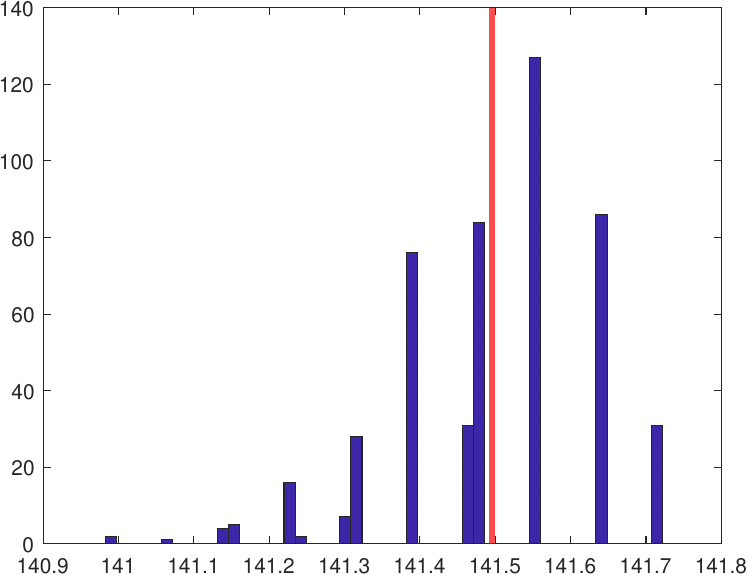} } 
\subfloat[CC, 10\% of nodes]{ \includegraphics[width = 0.33\textwidth]{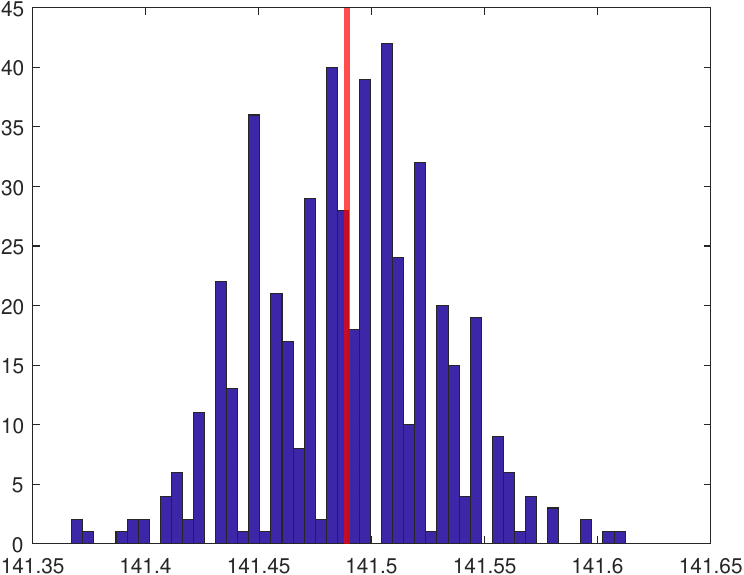} }
\caption{Distribution of $\delta$ statistic, 500 experiments, clusterable graphs} \label{Synthhists2}
\end{figure}

In Figure~\ref{Synthhists1}, we see that in all instances where the graph is not clusterable (ER, CM) the test statistic is distributed about a mean (red line) that is $\approx 0$. Meanwhile, in Figure~\ref{Synthhists2}, we observe that in all cases where the graph is indeed clusterable (SBM, CC), the statistic is distributed about a mean that is $\gg 0$.

Given these results on synthetic graphs with known structures, we determine that samples of $1\%$ of all nodes offer the optimal accuracy vs. size trade-off.

\subsection{Real world graph illustrations}
After determining the optimal sampling size of $1\%$ of nodes, we apply our test to two large real-world network with known community structure. Because of their large size, these networks highlight the value of our test. Here again, the test was repeated 500 times on each network and the number of rejections was reported. 
\begin{table}[H]
\centering
\caption{Null hypothesis rejection, real-world instances} \label{RW500}
\begin{tabular}{cccc}
\toprule
Graph  & Prop Sample & Num rejects & Prop reject \\
\midrule
Amazon   & 0.01      & 500    & 1     \\
Wiki    & 0.01        & 500   & 1        \\
\bottomrule
\end{tabular}
\end{table}

\begin{figure}[H]
\centering
\subfloat[Amazon, 1\% of nodes]{\includegraphics[width=0.4\textwidth]{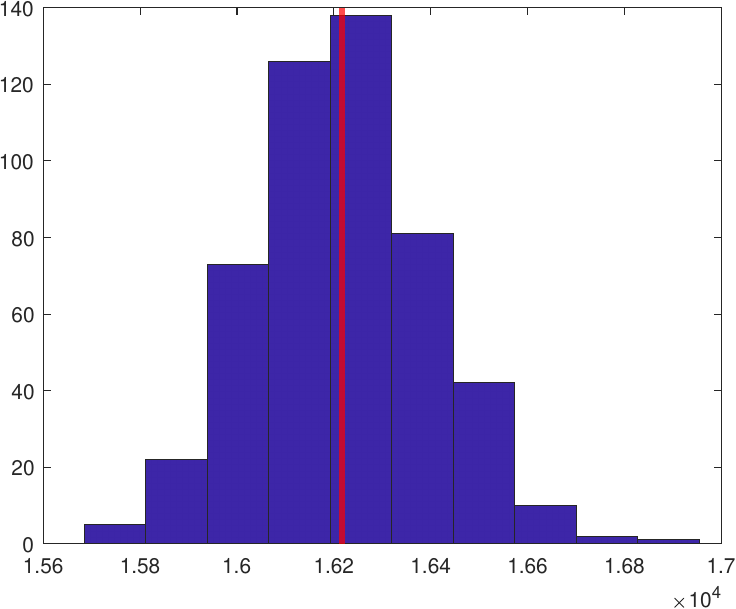}}
\subfloat[Wiki, 1\% of nodes]{\includegraphics[width=0.4\textwidth]{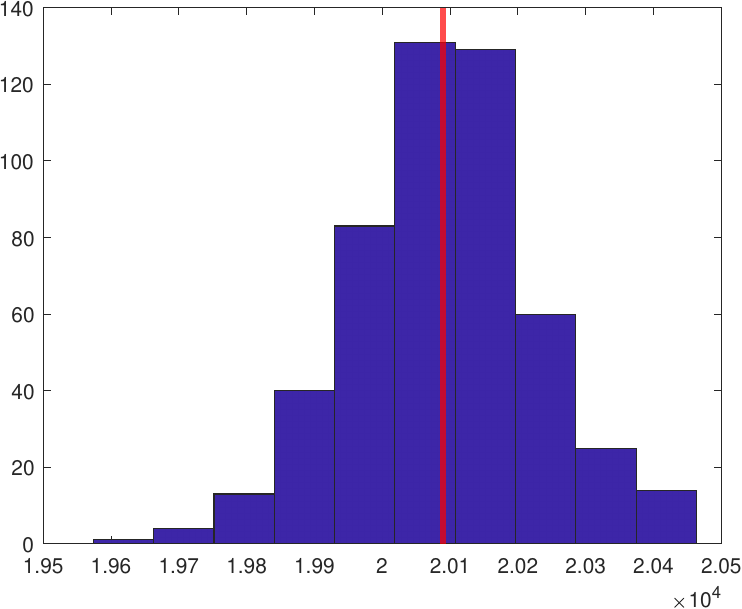}}
\caption{Real-world graphs with clusterable structure 1\%} \label{RWhists}	
\end{figure}

In both cases, we see that our test statistic is roughly symmetrically distributed about a mean that is $\gg 0$, indicating that the null is correctly rejected in all cases.

\subsection{Two graph test illustrations}
In Table~\ref{2S}, we show results from the two graph tests. We confirm the test's accuracy in determining the more clusterable graph in a pair. The known ``benchmarks'' (i.e., correct answer) for these tests is denoted as ``R'' (reject null) and ``NR'' (no rejection). Here, we remind the reader that the ER and CM graphs are stylized examples of unclusterable graphs. Meanwhile, the SBM graph is a stylized example of a graph with community structure. Furthermore, the Amazon and Wiki networks are known real-world instances of networks with community structure. In fact, our earlier tests also confirmed these claims from the literature. In Table~\ref{2S}, we also observe that test instances where the null should not be rejected have $0$ rejections, while cases where the null should be rejected are almst always rejected ($500/500$ and $498/500$).
\begin{table}[H]
\centering
\caption{Two sample testing results} \label{2S}
\begin{tabular}{cccc}
\toprule
More clustered & Less clustered & Benchmark & Num rejects \\
\midrule
ER             & Amazon         & NR        & 0           \\
ER             & SBM            & NR        & 0           \\
CC             & CM             & R         & 498         \\
Wiki           & CM             & R         & 500        \\
\bottomrule
\end{tabular}
\end{table}

\section{Analysis and discussion}
All tests have very good (almost perfect) accuracy, with samples of only $1\%$ of all nodes. Under three out of the four synthetic graph structures our test's conclusions are unaffected by sampling size. In both tests with real-world ground-truth community-structured networks, we obtain perfect results. These real-world networks highlight the value of our test, because of their large sizes. With just a small fraction of nodes (neighborhoods) sampled, our test correctly classified them as having  clustered structure.

Notably, however, the connected caveman graph (CC) graph highlights the test's limitations. Clearly, these limitations are related to the nature of the sample mean. Obviously, the sample mean of local densities is at the core of our test. As is well known, means are extremely sensitive to outliers. This sensitivity manifests itself clearly in the experiments on the CC graph. With small samples of local densities, a small number nodes with connections in two cliques (outliers) can deflate mean local density and hence the $\delta$ statistic. For example, in Figure~\ref{CCimg} sampling just two nodes $0$ and $1$ with neighborhoods $n_0$ and $n_1$ and local densities $\kappa_1$ and $\kappa_2$, we obtain the following mean local density (denoted $\bar{\kappa}$):
\begin{figure}[]
\centering
\includegraphics[width = 0.9\textwidth]{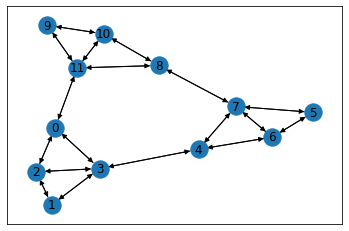}
\caption{Connected caveman digraph} \label{CCimg}
\end{figure}

\begin{eqnarray*}
\textit{First sampled neighborhood:} \\
n_0 &=& \{2,3,11\} \\
\vert e_0 \vert &=& 2 \\
\kappa_0 &=& \frac{2}{3 \times 2} = \frac{1}{3} \\
\textit{Second sampled neighborhood:} \\
n_1 &=& \{2,3\} \\
\vert e_1 \vert &=& 2 \\
\kappa_1 &=& \frac{2}{2 \times 1} = 1 \\
\textit{Aggregate (sample mean) level:} \\
&\Downarrow \\
\bar{\kappa} &=& \frac{1}{2} \left( 1 + \frac{1}{3} \right) = \frac{2}{3}
\end{eqnarray*}
Obviously, with larger random samples of nodes, the more probable nodes with no connections to other cliques will boost the mean local density and hence the $\delta$ statistic. This effect on the mean is also evident in Table~\ref{exp500}, where we see the proportion of (correct) rejections increase with sample size.

\begin{figure}[H]
\centering
\subfloat[CC local densities]{ \includegraphics[width = 0.4\textwidth]{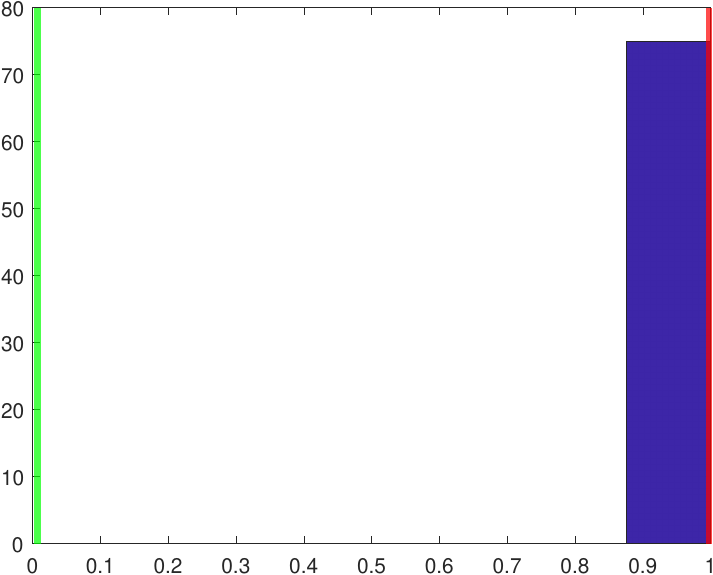} }
\subfloat[CM local densities]{ \includegraphics[width = 0.4\textwidth]{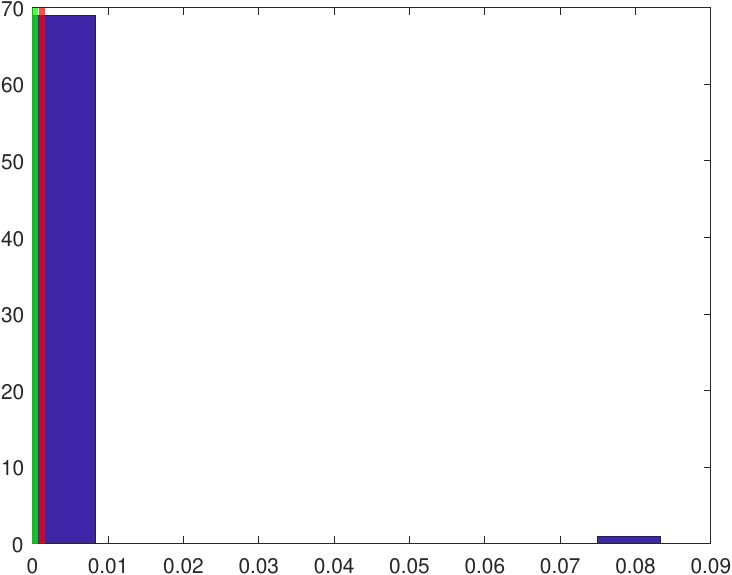} } 
\\
\subfloat[ER local densities]{ \includegraphics[width = 0.4\textwidth]{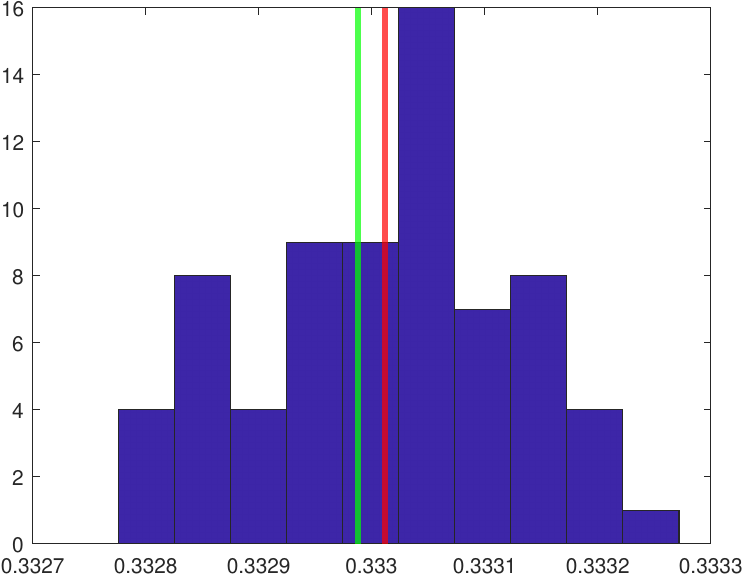} } 
\subfloat[SBM local densities]{ \includegraphics[width = 0.4\textwidth]{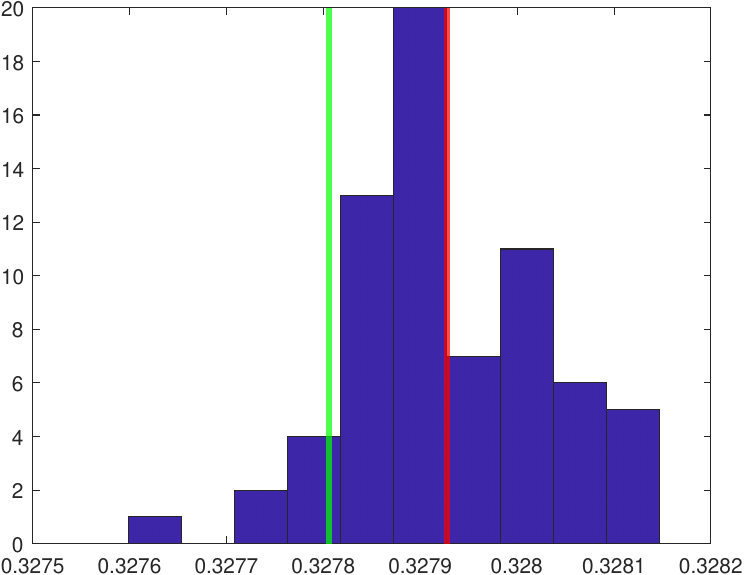} }
\\
\subfloat[Amazon local densities]{ \includegraphics[width = 0.4\textwidth]{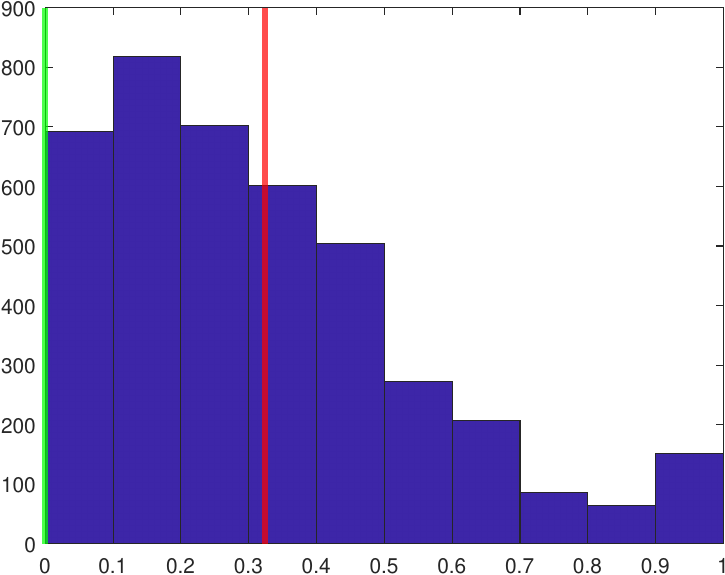} }
\subfloat[Wiki local densities]{ \includegraphics[width = 0.4\textwidth]{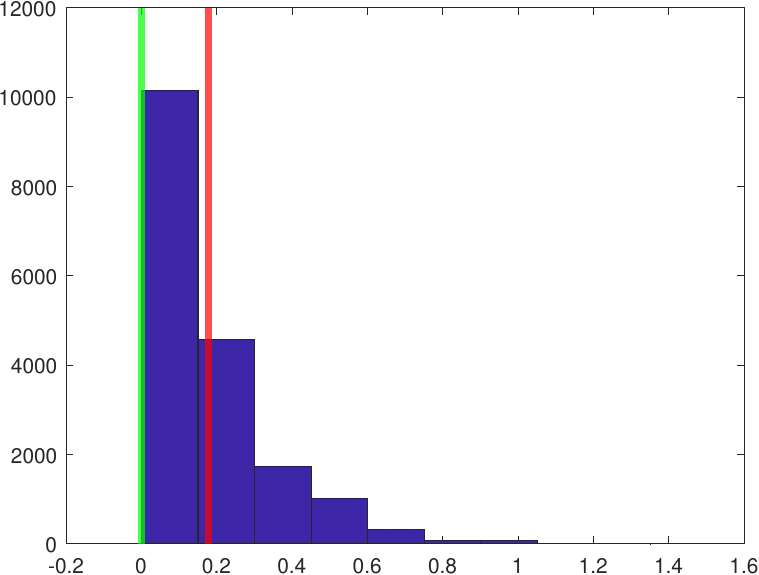} }
\caption{Distribution of local densities, single experiments with 1\% of nodes sampled} \label{hists1Exp}
\end{figure}

Figure~\ref{hists1Exp} illustrates the functioning and, of course, the limitation of using a mean-centric test. The histograms show the distributions of the local neighborhood densities. The red vertical line shows the mean of those local densities, the quantity which is tested. The green vertical line shows the graph's global density. In reviewing these figures, it is important to note that density is always contained in the interval $[0,1]$. We note that in cases where the graph is indeed clusterable (CC, SBM, Amazon, Wiki), the mean local density (red line) is higher than the graph's local density (green line). In contrast, in case where the graph is not clusterable (CM, ER) the mean local density (red line) appears minimally higher than the global density (green line), but this difference is not statistically different (higher).

On the topic of limitations, Figure~\ref{hists1Exp} shows that both the CC and CM graphs have highly concentrated non-symmetric distributions. These are instances where the mean does not offer a meaningful representative summary of the data. Arguably, both these graphs are stress test scenarios for our test. Fortunately, these synthetic graphs with their concentrated distributions are stylized models of extreme phenomena. Graphs with such concentrated local densities are, plausibly, rare. In fact, recent work has shown that this type of highly concentrated non-symmetric distribution, such as observed in scale-free networks, is rare in the real-world \cite{Broido2018}. 

Figures~\ref{Synthhists1},~\ref{Synthhists2},~\ref{RWhists} and~\ref{hists1Exp} also illustrate the CLT and its limitations very well. As just mentioned, densities, local or global, are always contained in the interval $[0,1]$. They may or may not be approximately Gaussian. Nevertheless, according to the CLT, their means and, consequently the $\delta$ statistic (normalized centered sample means) are indeed asymptotically Gaussian. We observe this approximately Gaussian distribution of the mean local densities computed with only $1\%$ of the neighborhoods (nodes), in all but the extreme cases of the CM and CC graphs. Regardless, even with these extreme cases, we do observe a clear trend toward a Gaussian distribution with larger samples, in the $\delta$ statistics (normalized centered means).

In closing, we argue that when applying the $\delta$ test, an examination of the distribution of local densities sampled should be part of the analysis workflow, in order to gain a better understanding of the connectivity structure and its suitability for clustering. As with any statistical examination, the mean cannot be the sole basis of any valid conclusion. 

\section{Conclusion}
We have demonstrated that the $\delta$ test can be extended to directed graphs. Future work will focus on stress test scenarios consisting of power law degree distributions. We will also examine a wider array of real-world graphs and study the entire density distributions, not just the means.

\section*{Acknowledgements}
The authors thank Andrei M. Raigorodskii and Liudmilla Prokhorenkova for their participation in the earlier developments of this test. 

\section*{Funding}
\begin{itemize}
\item The work of HD was funded through Project ECS 0000024 ``Ecosistema dell’innovazione - Rome Technopole'' financed by the EU NextGenerationEU plan through MUR Decree n. 1051 23.06.2022 PNRR Missione 4 Componente 2 Investimento 1.5 - CUP H33C22000420001.
\item The work of PM was financed by MITACS grant IT33832.
\end{itemize}

%\section*{Conflict of interests}
%The authors state they have no competing interests. 
%
%\section*{Author contributions}
%\begin{itemize}
%\item All authors reviewed this work and stand by it. 
%\item MRG conceived the idea of this extension to directed graphs.
%\item PM conceived the test and implemented it in Python.
%\item AS  reviewed the statistical methodology and designed and reviewed the experiments.
%\item HD and MRG conducted the empirical tests and compiled numerical results.
%\item CB and YL provided supervision and guidance throughout this work.
%\end{itemize}

\bibliography{biblio}

\end{document}